\pdfoutput=1

\documentclass[conference]{IEEEtran}
\usepackage{cite}

\usepackage{subfig} 
\usepackage{graphicx}
\usepackage{listings}
\usepackage{tabularx}
\usepackage{array}
\usepackage{paralist}

\usepackage{verbatim}
\usepackage{color}
\usepackage{xspace}

\newcommand\ourprotocol[0]{{GCBRR}\xspace}

\usepackage{algorithm}
\usepackage{algpseudocode}
\usepackage{amsmath}
\usepackage{amsfonts}
\usepackage{amssymb}
\usepackage{setspace}

\usepackage{fixltx2e}
\MakeRobust{\Call}

\algblockdefx[NAME]{function}{funcend}%
[2][]{\textbf{function} #1(#2)}%
{\textbf{end function}}

\algblockdefx[NAME]{template}{endtemplate}%
[2][]{\textbf{template} #1(#2)}%
{\textbf{end template}}

\algblockdefx[NAME]{protocol}{endprotocol}%
[2][]{\textbf{protocol} #1(#2)}%
{\textbf{end protocol}}

\ifCLASSINFOpdf
\else
\fi
\begin{document}
%

\title{Coordinated Broadcast-based Request-Reply and Group Management for Tightly-Coupled Wireless Systems}


\author{\IEEEauthorblockN{Manos Koutsoubelias and Spyros Lalis}
\IEEEauthorblockA{University of Thessaly \& CERTH\\
Volos, Greece \\
Email: \{emkouts,lalis\}@uth.gr}
}


%


\newcommand{\algrule}[1][.2pt]{\par\vskip.5\baselineskip\hrule height #1\par\vskip.5\baselineskip}

\maketitle

\begin{abstract}

As the domain of cyber-physical systems continues to grow, an increasing number of tightly-coupled distributed applications will be implemented on top of wireless networking technologies. Some of these applications, including collaborative robotic teams, work in a coordinated fashion, whereby a distinguished node takes control decisions and sends commands to other nodes, which in turn perform the requested action/operation and send back a reply/acknowledgment. The implementation of such interactions via reliable point-to-point flows may lead to a significant performance degradation due to collisions, especially when the system operates close to the capacity of the communication channel.  We propose a coordinated protocol which exploits the broadcast nature of the wireless medium in order to support this application-level interaction with a minimal number of message transmissions and predictable latency. The protocol also comes with group management functionality, allowing new processes to join and existing processes to leave the group in a controlled way. We evaluate a prototype implementation over WiFi, using a simulated setup as well as a physical testbed. Our results show that the proposed protocol can achieve significantly better performance compared to point-to- point approaches, and remains fully predictable and dependable even when operating close to the wireless channel capacity.
\end{abstract}


%
\IEEEpeerreviewmaketitle


 



\section{Introduction} 
\label{sec:introduction}

Many applications from the domain of cyber-physical systems \cite{Rajkumar:2010:CSN:1837274.1837461} implement a distributed control logic over a wireless medium. A typical example include teams of unmanned vehicles (UVs) that perform collaborative missions e.g., the aerial scanning of an area of interest with a certain flight formation \cite{Sujit2007}. Such applications, require tightly-coupled reliable interactions between the UVs over the wireless medium. Moreover, the aspect of control poses the requirement of predictable system responsiveness, which translates to predictable network interaction latencies.   

A popular implementation approach for these interactions is to adopt a coordinated design in the spirit of a group RPC \cite{ChangSeukLee1997} whereby a distinguished node takes control decisions and sends commands to other nodes, which in turn perform or schedule the requested action/operation and send back a reply/acknowledgment. Conventionally, such a communication pattern can be supported on top of multiple reliable point-to-point flows such as those provided from TCP. However this may lead to increased network traffic and higher latency. Furthermore, if performed in an un-coordinated way, it can lead to increased contention for the shared medium, resulting in unpredictable latencies. Both problems are particularly undesirable in a wireless setting, since they can lead to a significant performance degradation.

Taking a different approach, we propose a coordinated 1-N request-reply protocol, which achieves reliability, high throughput and predictable low latency. This is achieved by exploiting the broadcast nature of wireless communication in conjunction with a simple but effective scheme for avoiding contention during the transmission of replies back to the coordinator. The protocol also comes with integrated group management functions, allowing processes to join and leave the group at any point in time. Notably, our protocol can serve as a foundation for higher-level system support, such as the tightly-coupled UAV team programming approach presented in~\cite{Mottola:2014:TPD:2668332.2668353}.

The main contributions of the paper are: 
\begin{inparaenum}[(i)]
\item we present a coordinated protocol that supports request-reply interaction in dynamically changing wireless process groups;
\item we describe the protocol in detail, including a high-level algorithmic description and an informal proof for the key group management properties;
\item we evaluate an implementation of our protocol for a 802.11 WiFi network, using a simulated setup and a wireless testbed; 
\item we compare our protocol with TCP-based and simpler unicast-based approaches, showing that it achieves significantly better and predictable performance, especially when operating close to the channel capacity, for WiFi that favors unicast messaging. 
\end{inparaenum}
It is important to note that the proposed protocol is completely agnostic of the underlying network technology, and will work very well also on top of much simpler wireless technologies that do not feature an advanced medium-access-control (MAC) mechanism. 

The rest of the paper is structured as follows. Section~\ref{sec:api} introduces the application-level interface for a coordinated request-reply interaction within a process group. The protocol for supporting this API on top of a distributed system with broadcasting capability is described in detail in Section~\ref{sec:protocol}. Section~\ref{sec:perfprop} discusses the key performance properties and semantics of the supported request-reply interaction, while Section~\ref{sec:groupprop} gives informal proofs for the most important group management properties of the protocol. Section~\ref{sec:evaluation} presents a performance evaluation and comparison with different point-to-point approaches. 
Section~\ref{sec:related} provides an overview of related work. Finally, Section~\ref{sec:conclusions} concludes the paper and points to  directions for future work.

\section{Application interface} 
\label{sec:api}

We target distributed process groups, where a distinguished process acts as a \emph{coordinator} and issues requests to one or more \emph{ordinary} processes (nodes). In turn, the addressed ordinary processes perform the requested action or operation, and send back a reply. The reply serves as an application-level acknowledgment but may also contain additional information. 
Furthermore, new processes may join the group, and existing group members can leave the group, anytime. Processes that fail are also removed from the group. In case the coordinator fails or leaves the group, one of the ordinary processes takes over as the new coordinator. 

This functionality can be exposed to the application via the primitives of Table~\ref{tab:protocol_api}. 
The $RequestReply$ call and $ProcessRequest$ event embody the basic application-level request-reply interaction. The $Join$ and $Leave$ primitives let a process join and leave the group, respectively; it is up to the coordinator to check for and handle such processes, via $CheckJoins$ and $CheckLeaves$. Finally, the $BecameCoord$ event informs the application that it took over as the new coordinator, while $GroupChanged$ notifies it about any subsequent group changes. 

\begin{table}[!t]
\caption{Application-level primitives}
\label{tab:protocol_api}
\centering
\small
\begin{tabularx}{\columnwidth}{|X|}
\hline
\textbf{Init($pid,pids$):}
Initializes the protocol with identifier $pid$ for the local process. If $pids$ is empty, the local process must explicitly join the group via the $Join$ primitive. Else, $pids$ contains the members of the group, including $pid$.\\
\hline
\textbf{RequestReply($pids,data,procT$):}
Sends an application-level request to the set of processes $pids$ with payload $data$ and an estimated processing delay $procT$, and returns the set of replies produced. \\
\hline
\textbf{ProcessRequest($data$):}
Handle the next application-level request with payload $data$, and return the reply for it. \\
\hline
\textbf{Join():} 
Join the group. \\
\hline
\textbf{Leave():}
Leave the group. \\
\hline
\textbf{CheckJoins($joinT$):}
Wait $jointT$ for processes to join. \\
\hline
\textbf{CheckLeaves($leaveT$):}
Wait $leaveT$ for processes to leave. \\
\hline
\textbf{BecameCoord($pids$):}
The local process has coordinator status, for the group view which includes processes $pids$. \\
\hline
\textbf{GroupChanged($pids$):}
The current group view changed (due to a join/leave/failure), and includes processes $pids$. \\ 
\hline
\end{tabularx}
\end{table}

The motivation for introducing the 
$CheckJoins$ and $CheckLeaves$ primitives is to give the application more flexibility but also more control on the network traffic. For instance, the coordinator might let processes join or leave the group under certain application-specific conditions to which the underlying protocol is oblivious. Then, it is pointless (and costly) to keep the respective group management protocol functions active at all times. Also, some applications might be able to work without such group management functionality, in which case the corresponding protocol components could be dropped rather than be carried along as dead code; of course, the group could still change due to process failures.

Note that only the process that becomes the coordinator is notified accordingly, in order perform any application-level recovery actions, if needed, and resume the group's operation. Also, only the coordinator is notified about group changes so that it can adjust its operation.  
Note that the coordinator can easily forward such notifications to all other group members, under the control of the application, via the $RequestReply$ primitive. In a similar vein, it is up to the application to use this primitive in order to back-up critical coordinator state to other members, as needed. 
\section{The \ourprotocol protocol}
\label{sec:protocol}

To support the above API we propose a 1-N request-reply and group management protocol, which exploits the broadcast nature of the wireless medium to eliminate superfluous messages and avoid network contention. We refer to this protocol as the \emph{Coordinated Broadcast-based Group Request-Reply} protocol (or \ourprotocol).   

\subsection{System model and assumptions}

Let there be $N$ processes (nodes), with monotonically increasing identifiers, $p_i$, $1 \leq i \leq N$. 
The processes share the same broadcast domain and communicate via broadcast or unicast transmissions. 
Both types of transmissions are unreliable. Broadcasts are non-atomic and might arrive at only a subset of the processes. We assume bounded messaging delay: a message will arrive at its destination(s) within $MsgT$ after it was sent, or not at all. 

Processes may fail at any point in time. We assume synchronous and reliable fault detection: if $p_i$ fails then all correct processes will be notified about this failure within $FaultDetectT$. In case of multiple failures, the respective notifications may reach the correct processes in a different order. But a failure notification for $p_i$ cannot overtake the last message that was sent by $p_i$. Also, we assume that once a process leaves the group, it behaves like a failed process (in practice, this can be achieved by having the process explicitly reject messages coming from the group). 

Finally, a process that has failed or left the group might wish to join the group, again, at a later point in time. 
We assume that processes do not attempt to re-join earlier than $FaultDetect$ after they failed or left the group. This ensures that the messages sent by a process as part of the join procedure do not overtake the notification about its previous departure/failure. Alternatively, one can rely on process incarnation numbers.

\subsection{Protocol structure}

The \ourprotocol protocol is structured in a modular way. It consists of the: (i)~the basic \emph{application request-reply} protocol; (ii)~the \emph{join} protocol; (iii)~the \emph{leave} protocol; (iv)~the \emph{view-push} protocol; (v)~the \emph{fault handling and election} procedure. 
This makes it easier not only to describe but also to implement/test the protocol. Of course, these components are not entirely orthogonal to each other. But the interplay between them is limited and clearly defined. Also, it is straightforward to drop some of these components, if the corresponding functions are not required by the application.
In the following, we discuss the individual components of \ourprotocol.
Algorithms~\ref{alg:init_protocol}, \ref{alg:reqreply_protocol}, \ref{alg:join_protocol}, \ref{alg:leave_protocol}, \ref{alg:viewpush_protocol} and \ref{alg:fail_procedure} provide corresponding descriptions in the form of high-level code.  

\subsection{Protocol state and messages}

\begin{table}[!t]
\caption{Per-process protocol state}
\label{tab:protocol_state}
\centering
\small
\begin{tabular}{|l l|}
\hline
$state$ & Membership state \\
\hline
$myid$ & Identifier of the local process \\
\hline
$coordid$ & Identifier of the coordinator process  \\
\hline
$grp$ & Group view including ticket information \\
\hline
$seqno$ & Sequence number of last request sent/received \\
\hline
$rmask$ & Reply bitmask for the current request \\
\hline
$replies$ & Replies received from (ordinary) processes \\
\hline
$reply$ & Reply of local (ordinary) process  \\
\hline
\end{tabular}
\end{table}

\begin{table}[!t]
\caption{Protocol messages}
\label{tab:protocol_msgs}
\centering
\small
\begin{tabularx}{\columnwidth}{|X|}
\hline
\textbf{AREQ[$pid,k,rmask,data$]:} 
Application request from $pid$ (coord) with sequence number $k$ for processes in $rmask$ \\
\hline
\textbf{ARPL[$pid,k,data$]:}
Application reply from $pid$ for the request with sequence number $k$ \\
\hline
\hline
\textbf{JPOLL[$pid$]:}
Join poll from $pid$ (coord) \\
\hline
\textbf{JOIN[[$pid$]:}
Join request from $pid$, after a join poll \\
\hline
\hline
\textbf{LPOLL[]:}
Leave poll (from coord) \\
\hline
\textbf{LEFT[$pid$]:}
Leave notification from $pid$, after a leave poll \\
\hline
\hline
\textbf{VPUSH[$pid,grpc$]:}
View-push request from $pid$ (coord) for the new/updated group view $grpc$ \\
\hline
\textbf{VACK[$pid$]:}
View-push acknowledgment from $pid$\\
\hline
\end{tabularx}
\end{table}

The protocol state maintained locally by each process is summarized in Table~\ref{tab:protocol_state}. The messages types used in \ourprotocol for each of its components are shown in Table~\ref{tab:protocol_msgs}. 

The membership status of the local process is kept in variable $state$, and equals: $NORMAL$ if the process is a member of the group; $JOIN$ if the process is attempting to join the group (but is not yet a member of the group); $LEAVE$ if the process is attempting to leave the group (but is still a member of the group); $LEFT$ if the process has left the group. 

The current local group view, which contains the identifiers of the process that are considered to be members of the group, is kept in the $grp$ data structure. Importantly, $grp$ includes per-process ticket number information. The process with the smallest ticket number is the coordinator; its identifier is stored in $coordid$. The assignment of ticket numbers to newly joining processes is done by the coordinator. The group view also contains information in order to differentiate between old and newly added members.   

The sequence number of the last request sent/received (depending on whether the process is the coordinator or an ordinary process) is stored in $seqno$. The reply bitmask $rmask$ encodes the (ordinary) processes addressed in the current request, which should send back a reply: if $p_i$ should handle the request then the $i$th bit of $rmask$ is $1$ (else $0$). The reply bitmap also defines the order in which these processes should reply: if the $i$th bit is the $k$th non-zero bit of $rmask$ then $p_i$ must send its reply in the $k$th ``transmission slot''.
Ordinary processes store in $reply$ their reply to the last request received; this may have to be re-transmitted, if requested by the coordinator. The $replies$ structure is used by the coordinator to keep the replies that arrive from the ordinary processes. 

\subsection{Basic request-reply protocol}
When a new request-reply interaction starts, the sender/coordinator increases $seqno$, initializes $rmask$ to address the target processes, broadcasts a request message, and waits for the replies to arrive or a timer to expire. The timeout is set as a function of $MsgT$, the number of processes that are expected to reply, and the application request handling delay. Each reply that is received is added to the $replies$ structure, and $rmask$ is adjusted accordingly. When the timer fires and some processes have not replied, the request is re-transmitted. This is repeated until all target processes either reply or fail. The coordinator proceeds with the next application request once the previous one has been handled to completion. 

When a process receives a request, it checks $rmask$ to determine whether it is among the addressed processes. If so, it compares the sequence number $k$ with that of the last locally handled request $seqno$. If the request is new, it is handed over to the application, and the produce reply is stored in $reply$; else, the reply is already stored in $reply$. Replies are broadcasted, and as a result are received by other processes. The local process sends its reply following the order specified via $rmask$: if it is first in order, it does so as soon as the application produces the reply, else, right after it overhears the reply of the preceding process. To deal with message loss, a timer is set to expire in order to send the reply at the corresponding transmission slot.

\subsection{Join protocol}

The coordinator invites non-member processes to to join the group, by broadcasting a join poll message. It then waits a certain amount of time for corresponding requests. 
Each process that responds is added to the local group view, and is assigned the next ticket number. 

When the waiting time is over, the coordinator propagates the updated view to the group, via the view-push protocol (described in the sequel). In a first phase, the new view is disseminated to the old members (without the ones that are currently joining the group). In a second phase, the view is sent to each of the newly joining processes, in increasing ticket order.

A process wishing to join the group enters the $JOIN$ state and waits for the coordinator to start a join poll. It then sends a join request with its identifier (this is a unicast message). The process remains in this state until it receives a group view from the coordinator. In the meantime, it might receive additional join poll requests, to which it responds as above. Processes that are already members of the group ignore join polls.

\subsection{Leave protocol}

Similarly to the join protocol, the coordinator invites group members to declare their intention to leave, via a broadcast, and waits some time for any responses.

If an ordinary process wishes to leave, it enters the $LEAVE$ state, and waits for the coordinator to issue a leave poll (as long as the process remains in the $LEAVE$ state, it is expected to continue handling incoming requests as usual). It then announces its departure via a broadcast, and enters the $LEFT$ state. This state transition is unilateral, without any further communication. 
If the coordinator wishes to leave the group, it does the same (without having to wait for a leave poll).

When a process receives a departure announcement, it follows the same procedure as for process failures (explained in the sequel). Note that, since such announcements can be lost, the leave protocol is a merely \emph{best-effort} attempt to accelerate the removal of a process from the group view.  
If a departed process remains in the view of a member process, eventually, that process will receive a notification from the fault-detector.

\subsection{View-push protocol}
The goal of a view-push protocol is to commit a group view to a specified set of processes. It works along the lines of the request-reply protocol. In this particular case, the request carries the (updated) group view, and the target processes integrate it with their own local views, and send back an acknowledgment.

\subsection{Fault handling \& coordinator election}

When a process is notified about the failure (or departure) of $p$, it removes $p$ from its local group view. If $p$ is an ordinary process, the coordinator stops waiting for replies from it. If $p$ is the coordinator, the group member with the next smallest ticket number is elected as the new coordinator --- without any further communication among the remaining group members. Before resuming normal operation, the new coordinator performs a view-push, if its view contains new members. 

Note that the new coordinator can already be in the $LEAVE$ state. In this case, instead of taking over, it leaves the group following the leave protocol (see above), which in turn will trigger another election. 

\section{Performance properties and semantics}
\label{sec:perfprop}
\begin{algorithm}[hb] 
\caption{Protocol initialization}
\label{alg:init_protocol}
\setstretch{0.90}
\begin{algorithmic}[1]

\Function{Init}{$id,pids$}
  \State $myid \gets id$
  \State $seqno \gets 0$
  \State $grp \gets $ initGroupMembers($pids$)
  \If {$pids = \varnothing$}
    \State $state \gets LEFT$
    \State $coordid \gets 0$
  \Else 
    \State $state \gets NORMAL$
    \State $coordid \gets $ minTicket($grp$)
    \If {$myid = coordid$}
      \State {$\Call{BecameCoord}{$procs($grp$)}}
    \EndIf
  \EndIf
\EndFunction

\end{algorithmic}
\end{algorithm}
\setlength{\floatsep}{20pt plus 1.0pt minus 2.0pt}
\begin{algorithm}[!hb] 
\caption{Basic request-reply protocol}
\label{alg:reqreply_protocol}
\setstretch{0.90}
\begin{algorithmic}[1]

\Function{RequestReply}{$dsts,data,procT$}
  \State $replies \gets \varnothing$
  \State $seqno \gets seqno + 1$
  \State $rmask \gets $ setBits($dsts \ominus myid$)
  \While {$rmask \neq 0$}
    \State broadcast($AREQ[myid,seqno,rmask,data]$)
    \State $wt \gets MsgT + procT + $ bitsSet($rmask$)$*MsgT$
    \State \textbf{await}($rmask = 0,wt$)
  \EndWhile
  \State \Return ($replies$)
\EndFunction
\Statex

\function[OnRecv]{$AREQ[pid,k,rmask,data]$}
  \If{isBitSet($rmask,myid$)}
    \If{$k \neq seqno$}
      \State $seqno \gets k$
      \State $reply \gets \Call{ProcessRequest}{data}$
    \EndIf
    \State $myturn \gets $ false
    \State $wt \gets $ posBit($rmask,myid$)$*MsgT$
    \State await($myturn = $ true$,wt$)
    \State broadcast($ARPL[myid,seqno,reply]$)
  \EndIf
\funcend
\Statex

\function[OnRecv]{$ARPL[pid,k,data]$}
  \If {$k = seqno$}
    \If{$myid = coordid$}
      \State $replies \gets replies \oplus (pid,data)$
      \State $rmask \gets $ clearBit($rmask,pid$)
    \Else
      \State $myturn \gets $ nxtBit($rmask,pid,myid$)
    \EndIf
  \EndIf
\funcend
\end{algorithmic}
\end{algorithm}
\begin{algorithm} 
\caption{Join protocol}
\label{alg:join_protocol}
\setstretch{0.90}
\begin{algorithmic}[1]

\Function{Join}{} 
  \State $state \gets JOIN$
  \State \textbf{await}($state = NORMAL$)
\EndFunction
\Statex

\Function{CheckJoins}{$joinT$} 
  \State broadcast($JPOLL[myid]$)
  \State $wt \gets 2*MsgT + joinT$
  \State \textbf{sleep}($tw$)
  \If{new($grp$) $ \neq \varnothing$}
    \State ViewPush(old($grp$))
    \For {\textbf{each} $p \in $ sort(new($grp$))}
      \State ViewPush($\{p\}$)
    \EndFor
  \EndIf  
\EndFunction
\Statex

\function[OnRecv]{$JPOLL[pid]$}
  \If {$state = JOIN$}
    \State unicast($pid,JOIN[myid]$)
  \EndIf
\funcend
\Statex

\function[OnRecv]{$JOIN[pid]$}
  \State $t \gets $ nxtTicket($grp$)
  \State $grp \gets grp \oplus (pid,t)$
  \State $\Call{GroupUpdated}{grp}$
\funcend

\end{algorithmic}
\end{algorithm}

\begin{algorithm} 
\caption{Leave protocol}
\label{alg:leave_protocol}
\setstretch{0.90}
\begin{algorithmic}[1]

\Function{Leave}{} 
  \If{$mypid = coordid$} 
    \State $state \gets LEFT$
    \State broadcast($LEFT[myid]$) 
  \Else
    \State $state \gets LEAVE$
    \State \textbf{await}($state = LEFT$)
  \EndIf 
\EndFunction
\Statex

\Function{CheckLeaves}{$leaveT$} 
  \State broadcast($LPOLL[]$)
  \State $wt \gets 2*MsgT + leaveT$
  \State \textbf{sleep}($wt$)
\EndFunction
\Statex

\function[OnRecv]{$LPOLL[pid]$}
  \If {$state = LEAVE$}
    \State $state \gets LEFT$
    \State broadcast($LEFT[pid]$)
  \EndIf
\funcend
\Statex

\function[OnRecv]{$LEFT[pid]$}
  \State OnProcessFailure($pid$)
\funcend

\end{algorithmic}
\end{algorithm}

\begin{algorithm} 
\caption{View-push protocol}
\label{alg:viewpush_protocol}
\setstretch{0.90}
\begin{algorithmic}[1]

\function[ViewPush]{$pids$}
  \State $rmask \gets $ setBits($pids \ominus ypid$)
  \While {$rmask \neq 0$}
    \State broadcast($VPUSH[mypid,rmask,grp]$)
    \State $wt \gets MsgT + $ bitsSet($rmask$)$*MsgT$
    \State \textbf{await}($rmask = 0,wt$)
  \EndWhile
\funcend
\Statex

\function[OnRecv]{$VPUSH[pid,grpc]$} 
  \If {isBitSet($rmask,myid$)}
    \If{$pid \neq coordid$}
      \State $coordid \gets pid$
      \State $seqno \gets 0$
      \State $state \gets NORMAL$
    \EndIf  
    \For {\textbf{each} $(p,t) \in grpc$}
	\State $grp \gets grp \oplus (p,t)$
    \EndFor
    \State $myturn \gets $ false
    \State $wt \gets $ posBit($rmask,myid$)$*MsgT$
    \State await($myturn = $ true$,wt$)
    \State broadcast($VACK[mypid]$)
  \EndIf
\funcend
\Statex

\function[OnRecv]{$VACK[pid]$} 
  \If {$myid = coordid$}
    \State $rmask \gets $ clearBit($rmask,pid$)
  \Else 
    \State $myturn \gets $ nxtBit($rmask,pid,myid$)
  \EndIf
\funcend
  
\end{algorithmic}
\end{algorithm}

\begin{algorithm} 
\caption{Fault handling and coordinator election}
\label{alg:fail_procedure}
\setstretch{0.90}
\begin{algorithmic}[1]

\function[OnProcessFailure]{$pid$}
  \State $grp \gets grp \ominus (pid,*)$
  \If{$myid = coordid$}
    \State $rmask \gets $ clearBit($rmask,pid$)
    \State $\Call{GroupUpdated}{grp}$
  \ElsIf {$pid = coordid$}
    \State $coordid \gets $ minTicket($grp$)
    \If{$myid = coordid$}
      \If{$state = LEAVE$}
	\State $\Call{Leave}{}$
      \Else
	\State $\Call{BecameCoord}{grp}$
	\If{new($grp$) $ \neq \varnothing$}
	  \State ViewPush(old($grp$))
	  \For {\textbf{each} $p \in $ sort(new($grp$))}
	    \State ViewPush($\{p\}$)
	  \EndFor
	\EndIf
      \EndIf
    \EndIf
  \EndIf
\funcend

\end{algorithmic}
\end{algorithm}

The ``critical path'' of \ourprotocol in terms of performance is the basic request-reply protocol, which is designed to achieve several important properties: 
\begin{inparaenum}[(i)]
\item It is contention-free, avoiding concurrent transmissions from different processes.  
\item It incurs the minimum number of transmissions needed for a 1-to-N request-reply interaction: the request and each reply are transmitted just once (assuming no loss). 
\item It achieves minimal latency since ordinary processes send their replies as fast as possible, according to the schedule specified by the coordinator.
\item It is offers very predictable performance, and allows the system to operate close to the channel capacity. 
\item It enables robust message recovery in case of loss, by addressing only the processes that have not responded. 
\item The values for the message transmission delay and application-level request processing delay can be chosen generously, and affect performance only in case of message loss. 
\item Last but not least, no assumptions are made about the underlying medium-access-control (MAC) mechanism. Therefore \ourprotocol can work very well even on top of simple radios that do not have an intelligent MAC.
\end{inparaenum}

Admittedly, the join and leave protocols are subject to contention, if several processes wish to join and respectively leave the group at the same time. We consider this to be a rare occasion for most applications, and even then a simple back-off mechanism will greatly reduce the probability of collisions. Note, however, that the join/leave protocols do not interfere with the basic request-reply protocol; it is up to the coordinator to activate them when desired, via the respective primitives. 

In terms of application-level semantics, the request-reply interaction is synchronous, allowing the coordinator to drive the group in a tightly-coupled way, enforcing progress at all (addressed) processes at the same pace while also keeping track of their liveness.  
It is ``exactly once'' for processes that do not fail, and ``at most once'' for processes that fail. The application learns about such failures via the respective notifications; it can also infer a process failure by inspecting the returned replies. 
Notably, the interaction is ``non-atomic'', since a newly elected coordinator does not attempt to complete the last request-reply interaction that was initiated by the old coordinator. Atomicity does not seem to make sense here, because it is unclear how the new coordinator could handle, in a meaningful way, the replies for a request that was not issued by it. Of course, when the application is notified that it takes over as the new coordinator, it can perform any corrective actions it deems necessary.

\section{Group management properties}
\label{sec:groupprop}

The \ourprotocol protocol is designed for a group that has at least one correct process as a member at all times. 
If at some point all group members fail or leave, the group disappears and no longer exists. 
We also note that the (best-effort) leave protocol does not affect the essence of the group management functionality, and in the following consider process departures only due to failures.

Assume that group views contain entries $(p,t)$ where $p$ is a process identifier and $t$ the ticket assigned to it. It is easy to show that for any two correct group members $p_1$ and $p_2$ with local views $grp_1$ and $grp_2$, it holds that $(p,t_1) \in grp_1 \land (p,t_2) \in grp_2 \implies t_1 = t_2$. In other words, all members adopt the same ticket assignment (STA). Moreover, three additional important properties hold:
\begin{inparaenum}[(i)]
\item At any point in time, only one correct process $p$ can be elected as the coordinator (GM1).  
\item There is no deadlock, where the group is non-empty but it is not possible to elect a correct member process as the coordinator (GM2). 
\item The confirmation to a joining process that it became a member of the group is binding (GM3). 
\end{inparaenum}
Below we provide an informal proof sketch.  

\subsection{Uniqueness of the coordinator (GM1)}
From STA it directly follows that $grp_1 = grp_2 \implies $ \emph{minTicket}($grp_1$) $ = $ \emph{minTicket}($grp_2$). As a consequence GM1 holds for any two processes with the same view. However, during the transition periods caused by group dynamics, some group members might have different views. These cases are discussed below.

Let $(p,t) \in grp_1 \land (p,t) \notin grp_2$. If $p$ has failed, this does not affect GM1.  
It suffices to focus on views that differ only due to new processed joining the group.

Starting from a state where all group members have the same local view, assume that process $p_i, 1 \leq i \leq k$ join the group, and that the coordinator starts to push the new view that includes the respective entries $(p_i,t_i)$. Let the view push occur partially, reaching $p_1$ but not $p_2$, so that $(p_i,t_i) \in grp_1$ and $(p_i,t_i) \notin grp_2$.  
Since the view is first pushed to the old group members, $p_1$ cannot be a new member if $p_2$ is an old member. If both $p_1$ and $p_2$ are old members, due to the ticket assignment scheme it holds that $t_1,t_2 < t_i$, so the newly joined processes $p_i$ cannot affect the outcome of an election at $p_1$ or $p_2$. If $p_2$ is a new member, it is still in the JOIN state and cannot perform a coordinator election in the first place. Thus GM1 also holds during a group update phase.

\subsection{No deadlock (GM2)}
Given that GM1 holds, GM2 holds too, provided that if some correct process $p_2$ elects as the coordinator another correct process $p_1$  
then $p_1$ is certain to also consider itself as a member of the group. This is trivially so if $p_1$ is already a member when $p_2$ joins the group. It is somewhat less obvious if $p_1$ and $p_2$ join at the same time (respond to the same join poll). In this case, GM2 holds because the coordinator pushes the updated group view to each new member in increasing ticket order.

To see why, assume that the coordinator pushes the view $grp_c$ that includes entries for the joining members $(p_1,t_1)$ and $(p_1,t_2)$, with $t_1 < t_2$. Assume that the coordinator starts pushing the view in random order, but fails before completion. Also assume that all other old group members (which may have received $grp_c$) fail too. Finally, assume that $p_2$ has received $grp_c$ and considers itself as a group member, but $p_1$ has not. Then, $p_2$ will elect $p_1$ as the coordinator. However, $p_1$ remains in the $JOIN$ state, waiting for the next coordinator to initiate the next join poll, leading to a deadlock. 

\subsection{Join commitment (GM3)}

In order for GM3 to hold, it must be shown that once a joining process $p$ receives $grp_c$ with an entry $(p,t)$ for it, $p$ will never have to fall-back to the $JOIN$ state. Equivalently, it must be shown that if $p$ receives $grp_c$, every other process $p_i$ with $(p_i,t_i) \in grp_c \land t_i < t$, which could take over as a coordinator while $p$ is still alive, has already received $grp_c$. Indeed this holds due to the order in which the coordinator performs a view-push. 

Also, $p$ will not be addressed in the basic request-reply protocol, unless it considers itself as a member of the group (its state is $NORMAL$). This is guaranteed because the coordinator can proceed with the next request-reply interaction only after having successfully completed the corresponding view-push. It is also for this reason that a view-push is performed when a new coordinator takes over: to ensure that new members are aware of their membership status. This is not required if the view does not include any new group members.

\section{Evaluation} 
\label{sec:evaluation}

We have implemented \ourprotocol for a Linux-based platform, as a user-space library that resides on top of the operating system. Our implementation uses RAW sockets (bypassing the IP stack). The library offers a set of primitives in the spirit of the API in Table \ref{tab:protocol_api}. This section presents a performance evaluation over WiFi, and discusses the most important results.

\subsection{Experimental setup}
We test our implementation using two different setups, a simulated 802.11 network and a 802.11 testbed. 

\textbf{Simulated setup:} For the simulated setup we use the OpenNet simulator \cite{Chan2014}. This is an open-source simulation environment, which is build on top of Mininet \cite{Lantz:2010:NLR:1868447.1868466} and ns-3, combining their features to provide a realistic virtual wireless testbed infrastructure on a single PC. OpenNet uses Mininet to create a network of virtual Linux hosts, which can run conventional applications without any modification. Each virtual host features a virtual network interface which is internally connected via a TAP device to an ns-3 node. In turn, ns-3 provides the modeling of the 802.11 wireless channel between virtual hosts. To reflect the testbed setup, we configure ns-3 for the 802.11g protocol with a transmission rate of 1Mbps for both unicasts and broadcasts.

\textbf{Testbed setup:} For the real measurements we use the NITOS testbed~\cite{Nitos} featuring ICARUS~\cite{icarus} wireless nodes. These are PC-class machines with an Intel® Core™ i7-2600 Processor with 4 GB RAM and wireless Atheros 802.11a/b/g/n cards, running Linux. The testbed environment is protected from external traffic/interference. The nodes are placed in a grid, and can communicate with each other in 1 hop. The WiFi interface is accessed through the standard socket interface via \emph{ath9k} network driver, and is configured to operate with the 802.11g protocol in ad-hoc mode. To make a fair comparison among broadcast-based and unicast-based approaches, we fix the transmission rate to 1 Mbps for both. It would also be possible to transmit broadcast packets at the highest rate of the wireless interface by using the pseudo-broadcast transmission approach presented in~\cite{Katti2008}, but this would not contribute to the essence of our evaluation (the main point is for broadcasts and unicasts to be equally fast). The average round-trip time between two nodes is about 26 milliseconds for packets with a payload of 1500 bytes. 

The simulated setup is used to perform experiments across a wide range of parameters, without requiring physical nodes. We then use the testbed to test selected scenarios in order to verify the trends we observed via simulations. In both cases, each process (group member) is placed on a different virtual/physical node.

\subsection{Basic 1-N request-reply}

To evaluate the raw performance of the basic request-reply interaction of \ourprotocol, we use an application that issues dummy requests and produces dummy replies, without performing any processing. Request and reply messages are filled with dummy data, so all transmitted packets have the maximum payload size (1500 bytes). 

We compare \ourprotocol with four reliable point-to-point transports: TCP-SEQ, TCP-PAR, RUP-SEQ and RUP-PAR (RUP stands for reliable unicast protocol). In the TCP variants, the coordinator keeps a separate connection with each ordinary process, which is reused for all interactions. RUP variants use plain datagrams over RAW sockets, with a simple acknowledgment scheme for the re-transmission of requests. 
SEQ variants perform the request-reply interaction sequentially, one process at a time; this avoids contention but comes at increased latency. In the PAR variants, the coordinator first sends the request to all ordinary processes, and then waits for the replies; this can reduce the total latency but also introduces more contention. 

\begin{figure}[!t]
\centering
\begin{tabular}{c}
\includegraphics[width=0.80\columnwidth]{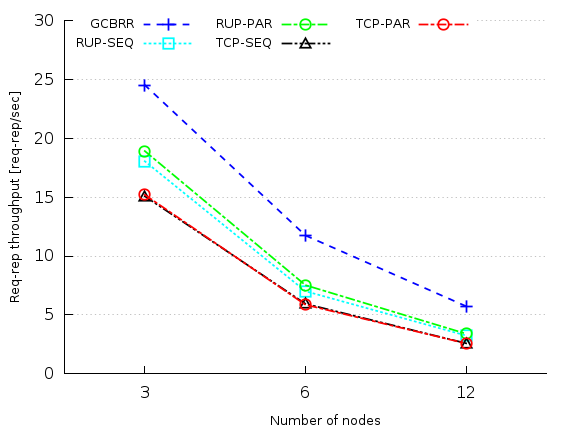} \\
\includegraphics[width=0.80\columnwidth]{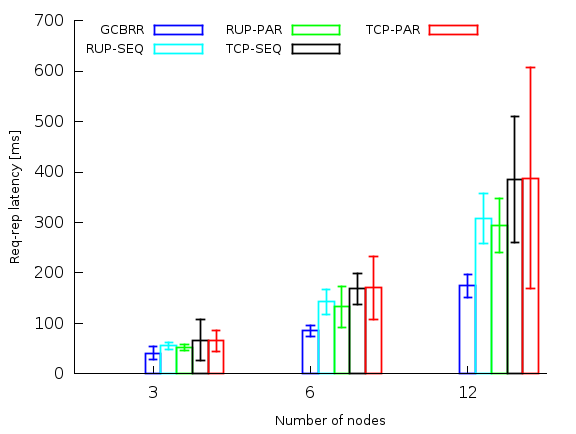} \\ 
\end{tabular}
\caption{1-to-N request-reply throughput (top), and latency (bottom), for different group sizes.}
\label{fig:maxout_1_N}
\end{figure}

We perform experiments for groups of different sizes: 3, 6 and 12 nodes. To stress the network, we let the coordinator issue 1000 request-reply interactions at full speed, and record the total throughput and latency of each interaction. The results are shown in Figure~\ref{fig:maxout_1_N}. 

As can be seen, \ourprotocol performs significantly better than all other variants, and the difference increases with the group size. For 3 nodes, \ourprotocol achieves x1.62 and x1.35 the throughput of TCP and RUP variants, going up to x2.22 and respectively x1.76 for 12 nodes. This is because \ourprotocol transmits fewer packets and scales better for a larger number of nodes. Note that for each unicast WiFi sends a MAC-level acknowledgement, which further increases the overhead of point-to-point variants. TCP achieves lower throughput than RUP due to the extra (TCP-level) acknowledgments, yet this difference becomes less pronounced for larger group sizes. Message loss is negligible in all cases, as the WiFi flow-control mechanism kicks in when the network is stressed.

The difference in the number of packet transmissions also reflects on the latency of the request-reply interaction. Again, \ourprotocol outperforms all other variants, especially in larger groups. For 12 nodes, its latency is only about $1/2$ and $2/3$ that of TCP and RUP, respectively. RUP variants are generally better than TCP. Note that RUP-PAR has a slightly lower latency than RUP-SEQ because of the concurrent transmission of replies back to the coordinator. This advantage does not show in the TCP variants, due to their higher network traffic. In fact, TCP-PAR has a much larger variance than TCP-SEQ. This is an effect of the increased contention: although there is no message loss, WiFi slows down transmissions internally.   

\subsection{One-way N-1 messages}

We also evaluate \ourprotocol regarding its ability to support a reverse, one-way message flow, from ordinary processes to the coordinator. In this case, the coordinator polls ordinary processes via an empty request, and receives their messages as replies. For comparison, we use TCP-PAR and RUP-PAR, where each ordinary process sends its messages to the coordinator over a TCP connection and a RAW socket, respectively. In RUP-PAR, messages are acknowledged explicitly following an alternating bit scheme. 
As an additional reference, we use the simplest possible best-effort approach where messages are sent via unreliable unicasts over a RAW socket, referred to as unreliable unicast protocol (UUP-PAR). 

\begin{figure}[!t]
\centering
\begin{tabular}{c}
\includegraphics[width=0.80\columnwidth]{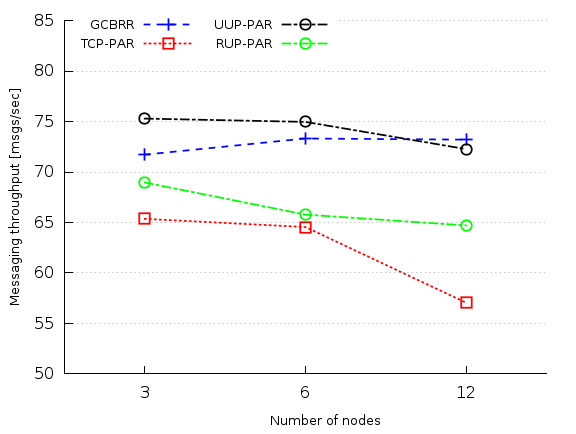} \\
\includegraphics[width=0.80\columnwidth]{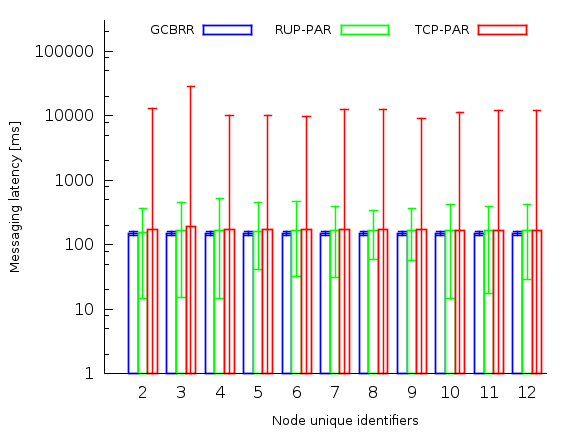} \\ 
\end{tabular}
\caption{Aggregate N-to-1 message throughput for different group sizes (top), and per-node messaging delay for an indicative run with 12 nodes (bottom; y-axis in log scale).}
\label{fig:maxout_N_1}
\end{figure}

We measure the maximum message throughput that can be achieved in a group of 3, 6 and 12 nodes, at full speed. In \ourprotocol, the coordinator polls the ordinary processes 1000 times, while in TCP-PAR, RUP-PAR and UUP-PAR we let each process send 1000 messages to the coordinator. The aggregate throughput is calculated at the coordinator, by recording the time between the arrival of the first and the last message. At each ordinary process, we record the time that elapses between two consecutive message transmissions, which reflects the messaging delay. The results are shown in Figure~\ref{fig:maxout_N_1}.

Despite the polling overhead, \ourprotocol achieves higher message throughput than TCP-PAR and RUP-PAR, even for smaller groups. The difference increases for larger groups, where TCP-PAR and RUP-PAR lead to increased contention, and the relative cost of polling decreases. At 12 nodes, the throughput of \ourprotocol is x1.28 and x1.13 that of TCP-PAR and RUP-PAR, respectively. In this case, \ourprotocol even outperforms UUP-PAR which also has a very significant message loss of about 7\%. Furthermore, unlike the RUP and TCP variants, \ourprotocol has a stable and predictable messaging delay. In fact, with TCP-PAR all nodes occasionally experience huge delays (the scale of the y-axis in the plot is logarithmic). Due to contention, TCP-PAR has some message loss, which leads to retransmissions and the activation of the TCP flow-control mechanism. In contrast, \ourprotocol and RUP-PAR did not need to perform any retransmissions. Also note that in TCP-PAR message sending sometimes appears to be instantaneous due to internal buffering.  

We wish to note that \ourprotocol, due to its polling approach, is not suitable for sporadic messaging from ordinary processes to the coordinator. In such scenarios, where the contention on the network is also low, it is better to use an uncoordinated approach.

\subsection{Group management}

Finally, we evaluate \ourprotocol in terms of the group management overhead. 

As a reference, we use a TCP-based approach where the coordinator keeps an open connection with each member process. A joining process opens a TCP connection to the coordinator when it receives a join poll (without sending a join request). It considers itself as a member once it successfully receives the group view over the connection. To leave the group, a process simply closes the connection to the coordinator (it does not send a leave announcement). If the coordinator wishes to leave the group, it closes the TCP connections to each group member. Ordinary processes detect that the connection was closed, elect the new coordinator and open a connection to it. The coordinator performs a view-push as in \ourprotocol, but in this case the group view is sent to each member over the respective TCP connection. As an implementation detail, we note that the coordinator uses a single thread for accepting new connections via a network event loop based on the $poll$ Linux system call, similar to the technique described in~\cite{Chandra01scalabilityof}. 

\begin{figure}[!t]
\centering
\begin{tabular}{c}
\includegraphics[width=0.80\columnwidth]{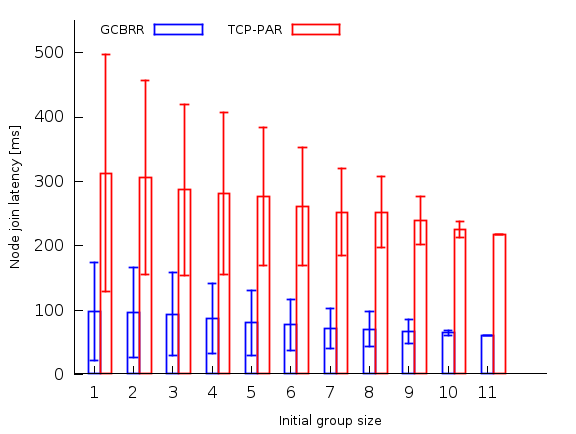} \\
\includegraphics[width=0.80\columnwidth]{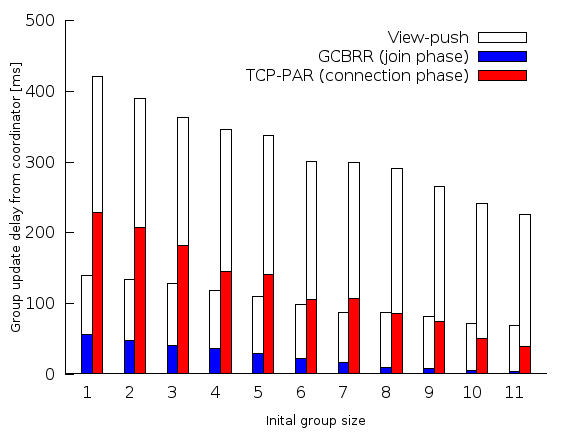} \\
\end{tabular}
\caption{Node join delay (top), and group update delay (bottom), when starting from initial group size $N$ and adding $12-N$ nodes in one shot.}
\label{fig:join}
\end{figure}

To chart the cost of the join operation as a function of the group size and number joining processes, we conduct a series of experiments. In each experiment, the group starts with a different initial size, and then more nodes join so that the group ends up with a total of 12 nodes. The new nodes join the group following a single join poll. At each joining node, we record the time between the reception of the poll message until the reception of the group view from the coordinator. We also measure the time it takes for the coordinator to receive all join/connection requests and push the updated view to all members. The results are shown in Figure~\ref{fig:join}.

With both approaches the join delay naturally decreases as fewer nodes try to join the group. However, the average node join delay and group update delay for \ourprotocol is only $1/3$ that of the TCP-based approach.
This is because the TCP variant has an extra connection setup overhead, which becomes significant when the number of joining nodes is large.  
Also, it sends the group view to each member separately, whereas \ourprotocol does this with a single transmission for old group members. 

The leave operation for ordinary group members is very fast, and the overhead negligible with both approaches. However, in case of a coordinator departure, in the TCP variant every ordinary process must open a connection to the new coordinator. The corresponding overhead follows the same trend as in the join experiments discussed above (it is not shown here for brevity). In contrast, the cost for such a coordinator switch in \ourprotocol is practically zero. Of course, in both approaches, the new coordinator may have to perform a view push. 
 
\section{Related Work} 
\label{sec:related}

The reliability issues of multicast communication patterns over a shared medium have been studied mainly in the lower level of the network stack. 
 
In \cite{Kuri1999} authors propose two reliable multicast protocols on top of IEEE 802.11 mechanisms.
Both protocols assume a base station in the center of a microcell and one single sender at a time.
The first protocol (LBP), is a leader based protocol while the other (PBP) is a probabilistic feedback-based 
protocol. In LBP, a sender claims the channel and sends a broadcast message to the receivers 
in the group where an elected leader responds with an acknowledgement. The rest of the receivers respond immediately with 
a negative acknowledgement in case of erroneous reception to destroy the acknowledgement of the leader causing a retransmission from the sender. 
In PBP a sender before the data transmission sends a multicast-RTS and waits to receive a CTS from each participant receiver. The receivers back-off with 
a certain probability based on the group size before the CTS transmission to avoid collisions. Both protocols increase broadcast
reliability, however they do not guarantee collision avoidance and require strict clock synchronization.

A similar approach with focus on fairness is followed in LM-ARF \cite{Choi2007} where an elected leader acknowledges 
mulitcast transmissions with the use of CTS/RTS frames. Moreover, LM-ARF performs an adaptation of the transmission rate based on the 
ARF \cite{Kamerman1997} scheme of 802.11 to avoid inefficient broadcast transmissions when network saturates to its capacity.
LM-ARF protocol inherits the time synchronization requirements from LBP and is closely coupled to the underlying MAC protocol.

The Broadcast Medium Window method (BMW)\cite{Tang2001} provides a reliable broadcast solution where the nodes maintain 
lists for their neighbors and for the sequence numbers of the missing data packets. Each sender performs a collision avoidance phase and sends an RTS to 
inquire the missing messages from its neighborhood. In turn, each neighbor responds with a CTS that contains the requested information and the sender 
performs the transmission of the data. Each neighbor updates the local protocol state by overhearing the CTS/data exchange. BMW provides a 
reliable solution for broadcast, but requires a large number of messages and creates contention phases equal to the number of neighbors.

Complementary to the BMW method is the Batch Mode Multicast MAC (BMMM) \cite{MinTeSun2002} which makes use of a new 
frame type called AK (Request for ACK) in order to eliminate the multiple contention phases of BMW to a single one. 
The RAK frame follows directly after the transmission of the RTS, instructing the receivers to send their CTS and ACKs in order.
BMMM follows an approach similar to our protocol with respect to the coordinated return of the replies, but involves  
more messages and abandons reliability when the network is stressed.

BSMA \cite{Tang2000} approach improves the broadcast reliability assuming that the underlying radio hardware 
supports the DS (direct sequence) capture capability in order to lock-on a strong signal in the presence of interference. 
The protocol utilizes the 802.11 RTS/CTS frames to eliminate collision avoidance and makes use of 
negative acknowledgments for prompting re-transmissions.The results show that throughput and reliability increases
over the traditional unreliable broadcast but the protocol relies strictly on 802.11 collision avoidance and requires specific radio capabilities. 
In the opposite, \ourprotocol is MAC-neutral while it is not based on any specific radio functionality.

The works above improve the physical broadcast reliability, however they do not have predicable performance which is a useful property for many latency sensitive applications. They also target general one-way multicast communication where the receiver merely sends a low-level acknowledgement, as opposed to \ourprotocol which works in an end-to-end fashion \cite{Saltzer:1984:EAS:357401.357402} and where replies may carry application-level information. 

\section{Conclusions} 
\label{sec:conclusions}

We propose a protocol for coordinated request-reply and group management in wireless systems, which eliminates contention and exploits the broadcast nature of the shared medium to minimize the number of message transmissions and latency. The proposed protocol does not rely on any advanced MAC features, and can work on top of different networking technologies.

Our evaluation over 802.11 WiFi for scenarios where the system operates close to the network capacity, shows that the proposed protocol achieves significantly higher throughput, lower latency and better predictability than unicast-based point-to-point approaches. These results are encouraging given that WiFi comes with a very effective flow-control mechanism and MAC-level support for unicasts. Note that, in case of message loss (e.g., due to external interference) the proposed protocol would perform even better compared to unicast-based approaches, as it requires fewer packet transmissions.

In the future we will experiment with simpler radios, and plan to investigate the usage of this protocol to support symmetrical process groups with N-to-N message flows.

\bibliographystyle{IEEEtran}
\bibliography{srds}

\end{document}